\documentclass[a4paper,preprintnumbers,showpacs,twocolumn,superscriptaddress,nofootinbib,amsmath,amssymb]{revtex4-1}
\usepackage[dvips]{graphics}
\usepackage[hypertex]{hyperref}
\usepackage{color,hyperref}
\usepackage{times}
\usepackage{mathrsfs}
\hypersetup{
  colorlinks=true,
  linkcolor=blue,
  citecolor=magenta,
  filecolor=magenta,
  urlcolor=blue
}

\def\be{\begin{equation}}
\def\ee{\end{equation}}
\def\bea{\begin{eqnarray}}
\def\eea{\end{eqnarray}}

\newcommand{\mc}{\mathcal}

\usepackage{enumitem}
\usepackage{graphicx,subfigure}
\usepackage{dcolumn}
\usepackage{bm}
\usepackage{color}

\begin{document}

\title{Dust collapse in 4D Einstein-Gauss-Bonnet gravity}

\author{Daniele Malafarina}
\email{daniele.malafarina@nu.edu.kz}
\affiliation{Department of Physics, Nazarbayev University, 53 Kabanbay Batyr, 010000 Nur-Sultan, Kazakhstan}

\author{Bobir Toshmatov}
\email{bobir.toshmatov@nu.edu.kz}
\affiliation{Department of Physics, Nazarbayev University, 53 Kabanbay Batyr, 010000 Nur-Sultan, Kazakhstan}
\affiliation{Ulugh Beg Astronomical Institute, Astronomicheskaya 33, Tashkent 100052, Uzbekistan}
\affiliation{Tashkent Institute of Irrigation and Agricultural
Mechanization Engineers, Kori Niyoziy 39, Tashkent 100000,
Uzbekistan}

\author{Naresh Dadhich}
\email{nkd@iucaa.in}
\affiliation{Inter-University Centre for Astronomy and Astrophysics, Post Bag 4, Ganeshkhind, Pune 411 007, India}

\begin{abstract}

We consider gravitational collapse in the recently proposed 4D limit of Einstein-Gauss-Bonnet gravity. We show that for collapse
of a sphere made of homogeneous dust the process is qualitatively
similar to the case of pure Einstein's gravity. The singularity
forms as the endstate of collapse and it is trapped behind the
horizon at all times. However, and differently from Einstein's
theory, as a consequence of the Gauss-Bonnet
term, the collapsing cloud reaches the singularity with zero
velocity, and  the time of formation of the singularity is delayed
with respect to the pure Einstein case.

\end{abstract}

\maketitle

\section{Introduction}

Gauss-Bonnet (GB) gravity \cite{lovelock} is quadratic order Lovelock theory while Einstein is the linear order and it has been studied for decades as one of the most interesting higher order theories of gravity in $D>4$ dimension. It also arises as low energy limit of heterotic superstring theory as the higher curvature correction to General Relativity (GR) \cite{gross,bento}. Note that Einstein is linear order while GB is a quadratic Lovelock theory, and Lovelock theorem states that any given order $N$ theory is non-vacuous only in $D>2N$; i.e., the GB term in the equations of motion gives contributions only in $D>4$.
However, it was recently proposed that a non trivial 4-dimensional limit of GB theory might exist, thus avoiding Lovelock's theorem, if one considers a rescaling of the GB coupling constant $\alpha$ to cancel the vanishing term $(D-4)$ arising from the variation of the GB action \cite{glavan}.


The validity of this 4-dimensional Einstein-Gauss-Bonnet (4D-EGB) theory is at present under debate in view of the fact that the rescaling proposed does not lead to a valid action for the $D=4$ limit of the theory (see for example \cite{Ai} and \cite{Gurses}). However, the rescaling of the GB coupling $\alpha$ works at the level of equations of motion. As a result one has a valid set of equations of motion for systems with certain symmetries that can be treated as an effective prescription for a 4-dimensional theory still to be determined.

However, it is certain that 4-dimensional solutions, such as spherically symmetric ones, can be constructed from the 4-dimensional action of the 4D-EGB obtained with the proposed redefinition of $\alpha$. Furthermore, it is possible to give such solutions a physical interpretation and investigate their properties, even in the eventual absence of a general theory, much in the same way as people investigate features of toy models in quantum-gravity even without a full quantum theory of the gravitational field. Thus it could be considered as an effective theory with a specific prescription. Comparison of the properties of such solutions with the corresponding ones in Einstein's gravity can then help shed some light on the validity of the underlying theory.

Due to the above mentioned importance of the solution, in a very short period various properties of the 4D-EGB solution were considered by several authors. For example, the properties of accretion disks around a static black hole were considered in \cite{liu} and \cite{li}, rotating black hole solutions were considered in \cite{ghosh}, thermodynamics was considered in \cite{HosseiniMansoori:2020yfj}, gravitational lensing and shadow was considered in \cite{wei} and \cite{kumar}, quasinormal modes were investigated in \cite{konoplya} and \cite{churilova}, grey-body factors and Hawking radiation were studied in \cite{Konoplya:2020cbv}. Furthermore, the extension of this approach to the 4D Einstein-Lovelock theory and stability of black holes in the 4D EGB and Einstein-Lovelock gravities were studied in \cite{Konoplya:2020qqh} and \cite{Konoplya:2020juj,Li:2020tlo,Casalino:2020kbt}, respectively.

In the present work we consider another spherical solution of great physical interest, namely the gravitational collapse of a spherical cloud of non interacting particles (i.e. dust). Gravitational collapse has been a cornerstone of black hole physics for decades as it provides the mechanism through which black holes and space-time singularities can form via a dynamical process starting from an initially regular configuration (see for example
\cite{Joshi-book} and \cite{JM}). Collapse in GB gravity was considered by several authors (see for example \cite{Maeda,jinghan,ghosh2,zhou,jgd}) and the final fate shows a non trivial dependence on the dimensionality of the space-time.

In this paper we specialize the framework for collapse to the 4D limit of the theory proposed in \cite{glavan}. We show that the qualitative behaviour of homogeneous dust collapse is similar to the one in Einstein's theory. The effect of the presence of the GB term can be seen in a delay in the co-moving time of formation of the singularity.

The paper is organized as follows: In section \ref{EGB} we review the properties of spherical collapse in EGB gravity. In section \ref{4d} we derive the equation of motion and the structure of singularity curve and apparent horizon for the 4-dimensional limit of the theory and compare with the corresponding results in pure Einstein's gravity. Finally, in section \ref{conc} we briefly outline the implications of our result for the theory.

Throughout the paper we make use of units where $G=c=1$ and absorb the term $\kappa=8\pi$ in the field equations into the definition of the energy-momentum.

\section{Dust collapse in Gauss-Bonnet gravity}\label{EGB}

A model for collapse in D-dimensional EGB theory is constructed from the action
\be
\mc{S}=\int d^Dx\sqrt{-g}\left(R+\tilde{\alpha}L_{GB}\right)+\mc{S}_{\rm matter} \; ,
\ee
were $R$ is the Ricci scalar which provides the general relativistic part of the action, $g$ is the determinant of the metric and where the GB part of the Lagrangian $L_{GB}$ is given by
\be
L_{GB}= R^2+R_{\lambda\mu\nu\xi}R^{\lambda\mu\nu\xi}-4R_{\mu\nu}R^{\mu\nu} \; .
\ee

Here we consider the line element for a spherical D-dimensional dynamical matter cloud in co-moving coordinates
\be \label{line1}
ds^2_-=-e^{2\Phi}dt^2+e^{2\Psi}dr^2+R^2d\Omega^2_{D-2} \; ,
\ee
with $\Phi(r,t)$, $\Psi(r,t)$ and $R(r,t)$ to be determined from the field equations
\be
G_{\mu\nu}=G_{\mu\nu}^{(E)}+G_{\mu\nu}^{(GB)}=T_{\mu\nu} \; ,
\ee
where $G_{\mu\nu}^{(E)}$ is the Einstein's tensor, $T_{\mu\nu}$ is the energy
momentum tensor and $G_{\mu\nu}^{(GB)}$ is the GB term given by
\bea
G_{\mu\nu}^{(GB)}&=&-\frac{\tilde{\alpha}}{2}g_{\mu\nu}L_{GB}+\\ \nonumber
&+&2\tilde{\alpha}\left(RR_{\mu\nu}-2R_{\mu\lambda}R^\lambda_\nu+\right. \\ \nonumber
&-&\left.2R_{\mu\lambda\nu\xi}R^{\lambda\xi}-R_{\mu\lambda\xi\sigma}R^{\sigma\xi\lambda}_\nu\right) \; .
\eea
For simplicity, we will restrict to the case of dust collapse and therefore,
we take $T_\mu^\nu={\rm diag}(-\rho, 0, 0, 0)$. From the Bianchi identities
we get $\Phi'=0$, where primed quantities refer to derivatives with respect
to $r$. With an opportune rescaling of the co-moving time $t$ we can then set $\Phi=0$.
The off diagonal component of the field equations is
\bea
G_r^t&=&(D-2)\frac{(\dot{\Psi}R'-\dot{R}')}{e^{2\Psi}R}+ \\ \nonumber
&+&2(D-2)\alpha\frac{(\dot{\Psi}R'-\dot{R}')(e^{2\Psi}+e^{2\Psi}\dot{R}^2-R'^2)}{e^{4\Psi}R^3}=0 \; ,
\eea
where dotted quantities refer to derivatives with respect to $t$ and we have
introduced $\alpha=(D-4)(D-3)\tilde{\alpha}$. From the above equation we get
two branches for collapse determined by two different forms for $e^{2\Psi}$.
One corresponds to taking
\be
2\alpha(e^{2\Psi}+e^{2\Psi}\dot{R}^2-R'^2)+e^{2\Psi}R^2=0 \; ,
\ee
and leads to
\be
e^{2\Psi}=\frac{2\alpha R'^2}{R^2+2\alpha(\dot{R}^2+1)} \; .
\ee
The other corresponds to taking
\be
\dot{\Psi}R'-\dot{R}'=0 \; ,
\ee
and leads to
\be \label{Gtr}
e^{2\Psi}= \frac{R'^2}{E} \; ,
\ee
where $E(r)$ is a free function of the radial coordinate, which is
related to the initial velocity profile of the collapsing cloud.
In the following we will focus on collapse satisfying equation
\eqref{Gtr} due to its similarities with the corresponding equation
in Einstein's gravity. The remaining non trivial field equations can
be written as
\bea \label{rho}
-G^t_t&=&\rho= \frac{(D-2)F'}{2R^{D-2}R'} \; ,\\ \label{p}
G^r_r&=&p_r=-\frac{(D-2)\dot{F}}{2R^{D-2}\dot{R}} =0 \; ,
\eea
where we have defined the mass function of the system $F$, which is
equivalent to the Misner-Sharp mass in the purely relativistic case \cite{misner}, as
\be\label{F}
F(r)=\alpha R^{D-5}(\dot{R}^2+1-E)^2+R^{D-3}(\dot{R}^2+1-E) \; .
\ee
Once equations \eqref{rho} and \eqref{p} are satisfied, the remaining
field equations, namely $G_{\theta_i}^{\theta_i}$ ($i=1...D-2$) are
automatically satisfied. Equation \eqref{F} can be written in the form
of the equation of motion for the system as
\be\label{eom}
\frac{F(r)}{R^{D-5}}=\alpha\dot{R}^4+[2\alpha(1-E)+R^2]\dot{R}^2+\alpha(1-E)^2+R^2(1-E) \; .
\ee
Notice that for $\alpha=0$ the equation reduces to the usual equation of
motion for dust collapse in GR. The above equation is a partial differential
equation of the fourth power in $\dot{R}$ and therefore, it will in general
admit four separate branches of solutions. In fact equation \eqref{eom} is
quadratic in $\dot{R}^2$ which for collapse ensuing from infinity with zero
velocity (i.e. $E=1$) has one positive and the other negative root. Obviously,
the positive root is the only physically meaningful one, which will give two
solutions corresponding to collapsing and expanding dust cloud. Also, the
collapsing positive solution can be matched with a vacuum geometry at the
boundary, and thus this is the branch on which we will focus in next section.

The quantity $F(r)$ acts as a quasi-local mass for the collapsing dust cloud
and can be understood as representing the amount of matter contained within
the co-moving radius $r$ at any time $t$. The fact that $F$ does not depend
on $t$ is a consequence of the choice of non interacting particles for the
matter content, i.e. a consequence of equation \eqref{p} which implies $\dot{F}=0$.
This implies that there is no inflow or outflow of matter through any shell $r$
and in particular through the boundary of the cloud and therefore the collapsing
interior can be matched to a vacuum exterior. Setting the co-moving boundary of
the collapsing cloud at $r=r_b$ we can then interpret $F(r_b)$ as related to the
total mass of the system, and therefore it should be related to the mass parameter
$M$ in the exterior vacuum solution. This is indeed the case if the matching is
performed with the vacuum solution discovered by Boulware and Deser \cite{boul}
that has line element
\be\label{ext}
ds^2_+=-H(S)dT^2+\frac{dS^2}{H(S)}+S^2d\Omega^2_{n-2} \; ,
\ee
with
\be\label{H}
H(S)=1+\frac{S^2}{2\alpha}\left(1\pm\sqrt{1+\frac{4\alpha M}{(D-2)S^{D-1}}}\right) \; .
\ee
Notice that there are two branches of solutions depending on the sign in front of the
square root, one referring to attractive and the other repulsive mass points and hence
we choose the branch with minus sign. The two branches have the same behaviour for small
$S$, namely they approach a De Sitter-like solution (one attractive while the other
repulsive), while they exhibit different behaviour asymptotically, with the branch with
minus sign having Schwarzschild-like behaviour for large $S$~\cite{dadhich, torii}. The
boundary radius $r=r_b$ in the interior corresponds to the collapsing boundary of the
cloud $R_b(t)=R(r_b,t)$, while the same boundary as seen from the exterior is given by
$S=S_b(T)$, with $T=T(t)$ on the boundary determined by the matching condition for
continuity of $g_{00}$. The continuity of the metric on the unit $(D-2)$-sphere implies that
\be \label{boundary}
R_b(t)=S_b(T(t)) \; .
\ee
The total $D$-dimensional ADM mass $M$ is then related to $F(r_b)$ through the continuity
condition for the equation of motion at the boundary (see \cite{Maeda}). In the simple case
of marginally bound collapse, given by $E=1$, this gives
\be
\dot{R}_b^2=-\frac{R_b^2}{2\alpha}\left(1\pm\sqrt{1+\frac{4\alpha F(r_b)}{R_b^{D-1}}}\right) \; ,
\ee
in the interior and
\be
S_b(T(t))=-\frac{S_b^2}{2\alpha}\left(1\pm\sqrt{1+\frac{8\alpha M}{(D-2)S_b^{D-1}}}\right) \; ,
\ee
in the exterior. The above equations, together with the boundary condition in equation
\eqref{boundary}, can be satisfied only if
\be
2M=(D-2)F(r_b) \; .
\ee
The above setup can be easily applied to the theory proposed in \cite{glavan} to investigate
the final outcome of homogeneous collapse.

\section{Dust collapse in 4D Gauss-Bonnet}\label{4d}

We will now restrict the attention to the special case of collapse in $D=4$ dimensions with
the rescaling of the GB coupling given by $\tilde{\alpha}\rightarrow \tilde{\alpha}/(D-4)$,
which in our case, setting $D=4$, corresponds to $\alpha=(D-3)\tilde{\alpha}=\tilde{\alpha}$.
It is important to point out that due to the spherical symmetry of the system, the
reparametrization of $\alpha$ proposed in \cite{glavan} gives valid field equations for this
model if the lower dimensional limit is taken suppressing the extra dimensions of the
$(D-2)$-sphere part of the metric. The line element \eqref{line1} in this case reads
\be\label{line1}
ds^2_-=-dt^2+\frac{R'^2}{E}dr^2+R^2d\Omega_2^2 \; ,
\ee
with $d\Omega_2^2$ the usual line-element on the unit 2-sphere and the derivation of the field
equations proceeds exactly as described in the previous section.

For the sake of clarity, we can now implement some scaling relying on the gauge freedom to set
the initial radius as $R(0,r)=r$, the arbitrariness of the function $E(r)$ and some physical
requirement for the mass function (see \cite{JM} for details). We thus set
\bea
R(r,t)&=&ra(r,t) \; ,\\
E(r)&=&1-r^2f(r) \; ,\\
F(r)&=&m(r)r^3 \; .
\eea
The case of homogeneous dust necessarily implies that $a(r,t)=a(t)$, $m(r)=m_0$ and $f(r)=k$ so
that the equation of motion \eqref{eom} reduces to
\be\label{eq-a-t}
\alpha(\dot{a}^2+k)^2+a^2(\dot{a}^2+k)=m_0a \; ,
\ee
which coincides with the equation of motion for the Oppenheimer-Snyder-Datt (OSD) collapse
\cite{OS,datt} when $\alpha=0$. Notice that with the above rescaling Einstein's equation for
$\rho(t)$, i.e. equation \eqref{rho}, becomes $\rho=3m_0/a^3$ which is finite at the onset of
collapse and leads to divergent density for $a\rightarrow 0$. For illustrative purposes, it is
useful to consider the simplest case of marginally bound collapse, which corresponds to $k=0$.
Then, equation \eqref{eq-a-t} reduces to
\be\label{eq-a-t2}
\alpha\dot{a}^4+a^2\dot{a}^2=m_0a \; ,
\ee
For $\alpha=0$ we retrieve the usual solution for OSD which, with the initial condition $a(0)=1$
takes the form $a(t)=(1-3\sqrt{m_0}t/2)^{2/3}$. Equation \eqref{eq-a-t2} is a quartic equation
and in general it has four roots of which only one describes collapse and matches with the black
hole exterior \eqref{ext}. Of the other three roots two are complex and one describes an expanding
cloud. The root of equation \eqref{eq-a-t2} describing collapse is given by
\be\label{eq-a-t3}
\dot{a}=-\frac{a}{\sqrt{2\alpha}}\sqrt{\sqrt{1+\frac{4\alpha m_0}{a^3}}-1} \; .
\ee
\begin{figure}[t]
\centering
\includegraphics[width=0.48\textwidth]{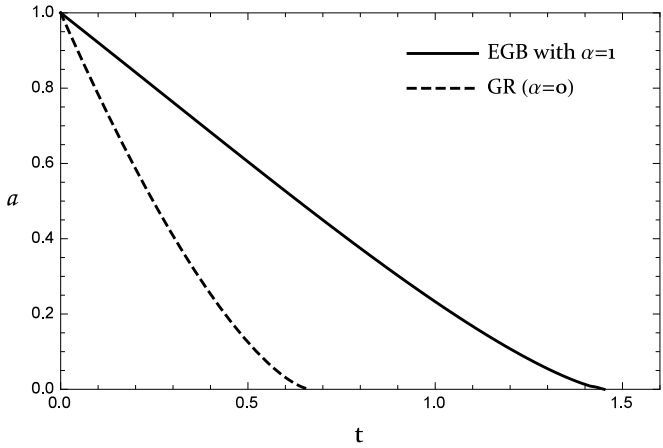}
\caption{\label{fig-a-t} The scale factor $a(t)$ for marginally bound dust collapse in Einstein's
theory (dashed line) and in the 4D-EGB theory (solid line). Notice that the effect of the GB term
on collapse is to slow the velocity of the scale factor and delay the time of formation of the
singularity. Here we choose the initial time $t=0$ in such a way that $a(0)=1$ for both cases.
Then, the singularity is reached faster in the purely relativistic case.}
\end{figure}
Notice that for this solution to be valid we must have $\alpha\neq 0$, so we can not retrieve the
OSD limit from the above equation. Analytical integration of equation \eqref{eq-a-t3} is
straightforward to obtain $t(a)$, which is a monotonic function of $a$ given by:
\be\label{t(a)}
t=\frac{2}{3}\sqrt{\alpha} \left(\arctan(z)-\frac{1}{z}\right)+c \; ,
\ee
where
\be
z=\frac{1}{\sqrt{2}} \sqrt{\sqrt{1+\frac{4\alpha m_0}{a^3}}-1}\; ,
\ee
and $c$ is an integration constant that is found by setting the initial condition as $a(0)=1$:
\be
c=\frac{2}{3}\sqrt{\alpha}\left(\frac{1}{z_0} -\arctan(z_0)\right)\; ,
\ee
with
\be
z_0=\sqrt{\frac{ \sqrt{1+4\alpha m_0}-1}{2}} \; .
\ee
Then the scale factor $a(t)$ is simply given by the inverse of equation \eqref{t(a)}. Comparison between
the behaviour of the solution of equation \eqref{eq-a-t3} and the OSD case is shown in figure~\ref{fig-a-t}.

\begin{figure}[t]
\centering
\includegraphics[width=0.48\textwidth]{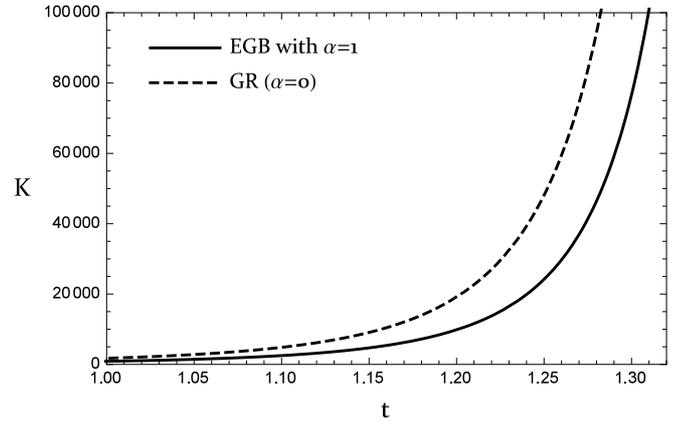}
\caption{\label{fig-curvature} Comparison between the Kretschmann scalar for collapse in Einstein's
gravity (dashed line) and in the 4D-EGB theory (solid line). Here we assume that the singularity occurs
at the same time $t_s$ for both scenarios (thus shifting the scale factor in the GR case). Then we can
see that, as a consequence of the GB term, $\mc{K}$ diverges faster in GR with respect to the 4D-EGB case.}
\end{figure}

\begin{figure}[t]
\centering
\includegraphics[width=0.48\textwidth]{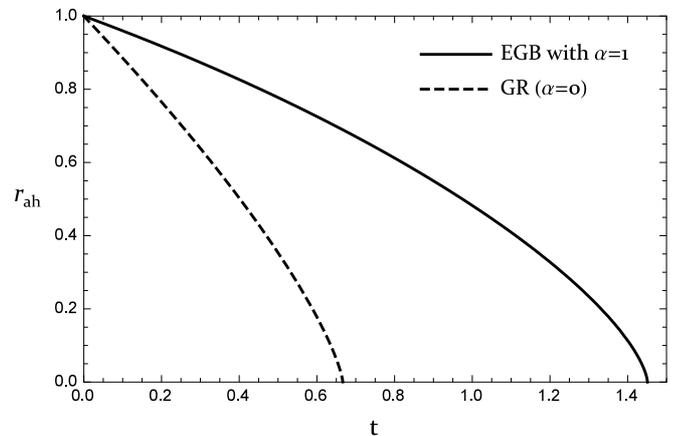}
\caption{\label{fig-a-h} Radius of apparent horizon curve as a function of time in pure Einstein's gravity
(dashed line) and in the newly proposed 4D-EGB theory (solid line). Again, assuming that collapse begins at
$t=0$ with $a(0)=1$, we see that the GB term affects the formation of trapped surfaces by delaying the
co-moving time at which each shell becomes trapped.}
\end{figure}

The Kretschmann scalar for the above scenario is given by
\be
\mc{K}=12\left(\frac{\ddot{a}^2}{a^2}+\frac{\dot{a}^4}{a^4}\right) \; ,
\ee
which diverges for $a\rightarrow 0$, thus showing that a curvature singularity forms as the endstate of
collapse, similarly to the purely relativistic case. However, as a consequence of the GB term the
singularity is ``weaker'' than in its relativistic counterpart. This can be seen from the fact that
the Kretschmann scalar diverges faster in the GR case, as shown in figure \ref{fig-curvature}.
From the above considerations we see that the singularity forms at the time $t_s$ for which $a(t_s)=0$.
Due to the homogeneous nature of collapse, all shells fall into the singularity at the same time $t_s$
and the singularity is space-like. In the present case we have
\be
t_s=\frac{2}{3} \sqrt{\alpha } \left(\frac{1}{z_0}-\arctan(z_0)\right)-\frac{\pi}{3}\sqrt{\alpha} \; .
\ee
We shall now investigate the behaviour of trapped surfaces for the marginally bound and homogeneous dust
collapse scenario. The apparent horizon in the interior is defined via the condition that the surface
$R(t,r)$ becomes null, i.e. $g^{\mu\nu}(\partial_\mu R)(\partial_\nu R)=0$. This corresponds to the requirement that
\be
1-\frac{F(r)}{R(t,r)}=1-\frac{r^2m_0}{a(t)}=0 \; ,
\ee
which implicitly defines the apparent horizon curve $r_{ah}(t)$ from
\be
r_{ah}(t)=\sqrt{\frac{a(t)}{m_0}}\; .
\ee
In figure \ref{fig-a-h} we show the evolution of the apparent horizon radius in the 4D-EGB theory as well
as in Einstein's theory. Similarly to what we have discussed for the scale factor, the main effect of the
GB term is to delay the formation of trapped surfaces.

\section{Conclusions}\label{conc}

We considered gravitational collapse of homogeneous dust in the newly proposed 4-dimensional limit of EGB
gravity. We showed that the scenario obtained within the theory proposed in~\cite{glavan} does not depart
significantly from the dynamical structure of dust collapse in Einstein's gravity. An appealing feature of
collapse in 4D-EGB theory is that the scale factor reaches the singularity with zero velocity thus making
the singularity at the end of collapse ``weaker" than in the corresponding relativistic case. It is known
that the metric function \eqref{H} remains regular as $r\to0$ as the approach to singularity occurs via de
Sitter-like space \cite{dadhich}. It is perhaps because of this reason that the singularity is weaker also
in collapse and the collapsing cloud reaches the singularity with zero velocity. Also, the 4D-EGB theory
allows for new branches of collapsing solutions that don't have a relativistic limit for $\alpha=0$ or at
large distances. These solutions will be investigated in future work. On the other hand, for the branch of
solution that reproduces to GR in the limit of $\alpha=0$ with asymptotically flat vacuum exterior, the
behaviour of the singularity and apparent horizon is qualitatively similar to the case of Einstein's theory
thus suggesting that black holes may form in 4D-EGB theory much in the same way as they do in GR.

It is well known that the introduction of inhomogeneities in dust collapse may alter dramatically the
formation of trapped surfaces leading to the appearance of naked singularities (see for example~\cite{Joshi}).
Therefore, it will be interesting to see if such conclusions remain valid for inhomogeneous dust collapse
in the 4D-EGB theory.

\section*{Acknowledgement}
This work was supported by Nazarbayev University Faculty Development Competitive Research Grant
No.~090118FD5348. BT acknowledges also the support from Uzbekistan Ministry for Innovative
Development Grants No.~VA-FA-F-2-008 and No.~MRB-AN-2019-29.

\end{document}